\documentclass[prl,twocolumn,preprintnumbers,superscriptaddress]{revtex4}

\usepackage{epsf}
\usepackage{natbib}
\usepackage{dcolumn}
\usepackage{bm}
\usepackage{ulem}
\usepackage{amsmath}
\usepackage{amssymb}
\usepackage{color}
\usepackage{latexsym} 
\usepackage{graphicx}
\usepackage{dcolumn}
\usepackage{bm}
\usepackage[
colorlinks=true,
citecolor=blue,
setpagesize=false]{hyperref}

\begin{document}


\title{Non-Fermi-liquid behavior and doping asymmetry in an organic Mott insulator interface}

\author{Yoshitaka Kawasugi} 
\affiliation{RIKEN, Wako, Saitama 351-0198, Japan}
\author{Kazuhiro Seki}
\affiliation{RIKEN, Wako, Saitama 351-0198, Japan}
\affiliation{SISSA - International School for Advanced Studies, Via Bonomea 265, 34136 Trieste, Italy}
\affiliation{RIKEN Center for Computational Science (R-CCS), Kobe, Hyogo 650-0047, Japan}
\author{Jiang Pu}
\affiliation{Department of Applied Physics, Nagoya University Furo-cho, Chikusa-ku, Nagoya, 464-8603, Japan}
\author{Taishi Takenobu}
\affiliation{Department of Applied Physics, Nagoya University Furo-cho, Chikusa-ku, Nagoya, 464-8603, Japan}
\author{Seiji Yunoki}
\affiliation{RIKEN, Wako, Saitama 351-0198, Japan}
\affiliation{RIKEN Center for Computational Science (R-CCS), Kobe, Hyogo 650-0047, Japan}
\affiliation{RIKEN Center for Emergent Matter Science (CEMS), Wako, Saitama 351-0198, Japan}
\author{Hiroshi M. Yamamoto} 
\affiliation{RIKEN, Wako, Saitama 351-0198, Japan}
\affiliation{Research Center of Integrative Molecular Systems (CIMoS), Institute for Molecular Science, National Institutes of Natural Sciences, Okazaki, Aichi 444-8585, Japan}
\author{Reizo Kato}
\affiliation{RIKEN, Wako, Saitama 351-0198, Japan}
 

\begin{abstract}
High-$T_{\rm C}$ superconductors show anomalous transport properties in their normal states, such as the bad-metal and pseudogap behaviors. 
To discuss their origins, it is important to speculate whether these behaviors are material-dependent or universal phenomena in the proximity of the Mott transition, by investigating similar but different material systems.
An organic Mott transistor is suitable for this purpose owing to the adjacency between the two-dimensional Mott insulating and superconducting states, simple electronic properties, and high doping/bandwidth tunability in the same sample.
Here we report the temperature dependence of the transport properties under electron and hole doping in an organic Mott electric-double-layer transistor.
At high temperatures, the bad-metal behavior widely appears except at half filling regardless of the doping polarity.
At lower temperatures, the pseudogap behavior is observed only under hole doping, while the Fermi-liquid-like behavior is observed under electron doping. 
The bad-metal behavior seems a universal high-energy scale phenomenon, while the pseudogap behavior is based on lower energy scale physics that can be influenced by details of the band structure.
\end{abstract}

\maketitle

\section*{Introduction}
Non-Fermi-liquid behaviors have been observed in the vicinity of two-dimensional Mott insulating and superconducting states in high-$T_{\rm C}$ cuprates \cite{Dagotto1994,Imada1998,Lee2006}.
The resistivity $\rho$ increases with temperature far above the Mott--Ioffe--Regel limit and shows $T$-linear behavior at sufficiently high temperatures (the bad metal behavior) \cite{Gurvitch1987,Takagi1992}.
The Hall coefficient $R_{\rm H}$ is strongly temperature-dependent despite the metallic-like transport \cite{Cheong1987,Ando2004}.
On the other hand, the cotangent of the Hall angle $\rho B/R_{\rm H}$ ($B$: magnetic field) is proportional to $T^{2}$ as in a Fermi liquid \cite{Chien1991}.
These anomalies have often been viewed as key unresolved signatures of strong correlation, which is probably relevant to the superconductivity.
To discuss their origins, it is important to investigate these behaviors in different systems which can be described with similar models, to speculate what behavior is material-dependent and what behavior universally appears in the proximity of the Mott transition.
Recently, we have developed electric-double-layer doping into an organic Mott insulator $\kappa$-(BEDT-TTF)$_{2}$Cu[N(CN)$_{2}$]Cl (hereafter $\kappa$-Cl), resulting in observation of ambipolar superconductivity surrounding the Mott insulating state in one and the same sample \cite{Kawasugi2018}. 
Therefore, our organic Mott electric-double-layer transistor (EDLT) would likely show the non-Fermi-liquid behaviors in the normal state.
In this study, we investigated the temperature dependence of surface resistivity, Hall coefficient, and Hall angle under electron and hole doping in an organic Mott EDLT, with varying doping across half filling in a single sample.
At high temperatures, the bad metal (BM) behavior widely appears except at half filling regardless of the doping polarity.
At lower temperatures, the transport properties are highly doing-asymmetric; the pseudogap behavior is observed under hole doping while the Fermi-liquid-like behavior is observed under electron doping. 
These results imply that the former behavior is a universal high-energy characteristic of the Mott transition, and the latter is a lattice- (band-structure-) dependent phenomenon.

$\kappa$-Cl is a two-dimensional organic Mott insulator, which comprises alternating layers of conducting BEDT-TTF$^{+0.5}$ radical cations and insulating Cu[N(CN)$_{2}$]Cl counter anions \cite{Williams1990}. 
The conducting BEDT-TTF molecules are strongly dimerized and can be modeled as a single-band Hubbard model on an anisotropic triangular lattice: $t'/t$ = $-$0.44 \cite{Kandpal2009}, where $t$ is the nearest-neighbor hopping and $t'$ is the next-nearest-neighbor hopping. 
The sign of $t'/t$ is negative and thus the van Hove singularity lies below the Fermi energy (hole-doped side) similarly to the high-$T_{\rm C}$ cuprates, although $t'$ exists only for one diagonal of the dimer sites and accordingly the Fermi surface is elliptical (Fig.~1).
The BM behaviors are common in normal states of doped cuprates, but field-effect-transistors (FETs) made of $\kappa$-Cl only showed insulator-to-metal crossover (IMC) and did not show any BM behavior in our previous experiments probably because the device conductance comprised both surface and bulk contributions, the latter of which followed only activation-type temperature dependence \cite{Sato2017}. 
Despite such a difficulty in separating the surface conductance from total one, it is important to reveal whether the BM behaviors exist in doped organic Mott insulator.
Comparison between doped cuprates and $\kappa$-BEDT-TTF EDLT, in terms of BM behaviors in the normal states, can address such an issue (Table~I).

By using the EDLT method, it is possible to define the charge neutrality point, or precisely half-filled Mott insulating state, to evaluate the surface transport properties (see Methods section). 
In this paper, we show the temperature dependence of surface resistivity $\rho _{\rm s}$ and surface Hall coefficient $R_{\rm Hs}$ in the $\kappa$-Cl EDLT up to 200 K, based on a simple two-fluid (doped surface and undoped bulk) analysis. 
Note that the strength of correlation ($U/t$) of $\kappa$-Cl in the present study is larger than those in Ref. \cite{Kawasugi2018} because we employed substrates with less coefficient of thermal expansion, to increase the ratio of surface conductance to bulk conductance. 
Therefore, the superconductivity does not appear in the present study.
In the following, experimental data taken from two samples will be shown; namely, the resistivity data are obtained from sample \#1 and the magneto-transport data are obtained from sample \#2. 
The electron and hole doping concentration is changed up to approximately 20\% according to the charge displacement current measurements in our previous work \cite{Kawasugi2018}.

\section*{Methods}
\subsection*{Sample preparation and transport measurement}
Thin single crystals of $\kappa$-Cl were electrochemically synthesized and laminated on a polyethylene naphthalate (PEN) substrate with patterned Au electrodes (see Fig. S1 in the Supplemental Material \cite{SupplementalMaterial}).
The thicknesses of the crystals measured using a step profiler were 31.5 nm for sample \#1 and 81.5 nm for sample \#2.
Following the laser shaping of the single crystals, we fabricated the EDLT devices by mounting a droplet of the ionic liquid 1-ethyl-3-methylimidazolium 2-(2-methoxyethoxy) ethylsulfate on a Hall-bar-shaped thin single crystal of $\kappa$-Cl and a Au side gate electrode.
A 1.2-$\mu$m-thick PEN film was then covered for the thinning of the droplet in order to reduce thermal stress at low temperatures.
The temperature and magnetic field were controlled using a Physical Property Measurement System (Quantum Design) at the sweep rates of 2 K/min ($T>20K$), 0.5 K/min ($T<20$ K), and 150 Oe/s.
Throughout the measurements, gate voltage was varied at 220 K because the ionic liquid is solidified at lower temperatures.
We measured the transport properties along the $a$- and $c$-axes, which were determined by the shape of the crystal and the sign of the Seebeck effect.
In the Hall measurements, the magnetic field was swept in the range of $\pm$8 Tesla at a constant temperature, and the forward and backward data were averaged to eliminate small linear voltage drifts.

In the EDLT, only the surface of the sample is doped.
However, the nondoped region of the sample is also conductive at moderate temperatures ($T\gtrapprox $50 K) in $\kappa$-Cl.
Assuming two parallel conducting layers, we derived the surface resistivity $\rho _{\rm s}$ and the surface Hall coefficient $R_{\rm Hs}$ from
\begin{equation}
\rho _{\rm s} = \left(\frac{1}{\rho}-\frac{1}{\rho _{\rm b}}\right)^{-1},
\end{equation}
\begin{equation}
R_{\rm Hs} = \rho _{\rm s}^{2} \left( \frac{R_{\rm H}}{\rho ^{2}}-\frac{R_{\rm Hb}}{\rho _{\rm b}^{2}}\right),
\end{equation}
where the suffixes b and s denote the bulk and surface, respectively. 
Here, we employed $\rho \times N/(N-1)$ and $R_{\rm H} \times N/(N-1)$ ($N$: number of conducting layers, $N=$21 and 54 in samples \#1 and \#2, respectively) at the charge neutrality point as $\rho _{\rm b}$ and $R_{\rm Hb}$, respectively (see Fig. S2 in the Supplemental Material \cite{SupplementalMaterial}).
The error bars in Fig. 3 were calculated from the standard deviation of the Hall resistance vs magnetic field plots, and the standard error propagation formula.
The surface to bulk conductivity ratios in sample \#1 were approximately 25\%, 19\%, 16\%, 24\% at $V_{\rm G}=-0.6, -0.3, +0.3, +0.6$ V, respectively.

The variational-cluster-approximation calculations for Fig. 1(d) and cluster-perturbation-theory calculations for Fig. 3(c) were performed for the Hubbard model on an anisotropic triangular lattice ($t'/t = -0.44$, $U/t = 5.5$, and $t = 65$ meV~\cite{Kandpal2009}) using a 12-site cluster (an antiferromagnetic insulating state at half filling is assumed in the former case).
Details of the calculations are described in our previous paper \cite{Kawasugi2016, Kawasugi2018}.

\section*{Results}
\textbf{Temperature dependence of resistivity} 
Figure 2(a) shows the temperature ($T$) dependence of the surface resistivity $\rho _{\rm s}$ under electron doping.
At high temperatures, metallic-like conduction (d$\rho _{\rm s}$/d$T>$0) above the Mott--Ioffe--Regel limit $\rho _{\rm MIR}$ is observed in a wide range of doping level.
Although the resistivity is not $T$-linear (between linear and quadratic) in this temperature range, this is a BM behavior in the sense that the mean-free path of carriers is shorter than the site distance.
The power-law exponent $\alpha$ estimated from $\rho _{\rm s}=\rho _{\rm 0}+AT^{\alpha}$ ($\rho _{\rm 0}$: residual resistivity, $A$: an arbitrary prefactor) is approximately 1.5 between 50 and 200 K at 0.6 V, although the exponent is not very accurate because the effect of thermal contraction (variations of the cell parameters \cite{Muller2002}) is not considered here.
At 20--50 K, this sample exhibits an IMC at gate voltage $V_{\rm G}$=0.34 V, namely the sign of d$\rho _{\rm s}$/d$T$ changes by the gate voltage.
The resistivity at 0.34 V is close to $h/e^{2}$, which corresponds to $\rho _{\rm MIR}$ for two-dimensional metal with an isotropic Fermi surface \cite{Thouless1975,Ahadi2017}.
By further electron doping, the temperature dependence of $\rho _{\rm s}$ approaches that of a Fermi liquid at low temperatures.
As shown in Fig. 2(c), $\rho _{\rm s}$ in the highly electron-doped states are almost quadratic in temperature below $\sim$20 K, although they slightly deviates from the $T^{2}$ line in the lowest temperature region because of the weak localization effect.

Under hole doping, the BM behavior is found at high temperatures as in the case of electron doping [Fig. 2(b)].
Although an accurate estimation of the power-low exponent is difficult owing to the resistivity upturn, the temperature dependence of $\rho _{\rm s}$ appears more $T$-linear than in the case of electron doping at high temperatures [Fig. 2(d)].
This is consistent with the $T$-linear resistivity in the hole-doped compound $\kappa$-(BEDT-TTF)$_{4}$Hg$_{2.89}$Br$_{8}$ at high temperatures \cite{Taniguchi2007,Oike2015}.
The gate voltages where the BM behavior occurs ($|V_{\rm G}|\sim $0.1 V) correspond to a few percent doping according to the Hall effect.
On the other hand, no IMC with varying $V_{\rm G}$ is observed at low temperatures.

Figure 2(e) shows color plots of more detailed $\rho _{\rm s}$ data under electron doping.
The result of calculations based on dynamical mean-field theory (DMFT) \cite{Dobro2015} is also shown as a reference.
The results of our experiment and the DMFT calculations are qualitatively similar at high temperatures, but somehow different at low temperatures: the high-resistance region in the experiment gradually expands with decreasing temperature because of Anderson localization.
As a result, the $\rho _{\rm MIR}$ line is shifted to a higher doping level and becomes almost vertical in the temperature range 20--50 K.
The IMC at $V_{\rm G}$=0.34 V is therefore related not only to the electron correlation but also to the surface potential disorder.

\textbf{Temperature dependence of Hall coefficient and Hall angle}
In a Fermi liquid with a single type of carrier, $1/eR_{\rm H}$ denotes the carrier density corresponding to the volume enclosed by the Fermi surface (FS) \cite{Luttinger1960}.
If the doping-driven Mott transition restores the noninteracting FS topology, $1/eR_{\rm H}$ coincides with the intrinsic carrier density, 1$-\delta$ hole per site where $\delta$ is the electron doping concentration.
 This is applicable to the case of surface electron doping.
$1/eR_{\rm Hs}$ is close to the half-filled carrier density (1 hole per site) and almost linearly decreases with $V_{\rm G}$.
By contrast, $1/eR_{\rm Hs}$ under hole doping is several times less than 1$-\delta$ at low temperature as previously reported \cite{Kawasugi2016}.
This is attributed to the emergence of an anisotropic pseudogap near the saddle points of the band dispersion that cause the van Hove singularity (the saddle points locate at the same wave vectors as those of the cuprates, ($\pm \pi$, 0) and (0, $\pm \pi$) in the single-band model as shown in Fig. 1) \cite{Kawasugi2016}.
The noninteracting ellipsoidal FS is truncated and $1/eR_{\rm Hs}$ reflects the volume of the lenslike hole pocket \cite{Oshima1988}.

The temperature dependences of $R_{\rm Hs}$ in Fig. 3(a) displays these features described above. 
Moreover, our experiment finds that $R_{\rm Hs}$ is almost temperature-independent under electron doping, while it significantly decreases with temperature under hole doping [and appears proportional to 1/$T$ above $\sim$50 K, as shown in Fig. 3(b)], as if the pseudogap diminishes and the noninteracting-like FS starts to contribute to $R_{\rm Hs}$ with increasing temperature.
Our finite-temperature cluster-perturbation-theory (CPT) calculations~\cite{Seki2018} also support the diminishment of the pseudogap with temperature [Fig. 3(c)].

Figure 3(d) shows the surface Hall angle at 1 Tesla $\rho_{\rm s} /R_{\rm Hs}$ (=inverse Hall mobility) vs $T^{2}$.
In the hole-doped states, $\rho_{\rm s} /R_{\rm Hs}$ is nearly quadratic in temperature with nonzero intersections.
The slope does not depend much on $V_{\rm G}$, whereas the $y$-axis intersection, i.e., $\rho_{\rm s} /R_{\rm Hs}$ at $T=0$, decreases with doping.
By contrast, the quadratic relation is less applicable in the electron-doped states.
Although the quadratic relation is suggested at low temperatures, in agreement with the Fermi liquid behavior [Fig. 2(c)], the plots bend downward at approximately 100 K and the slope becomes more $V_{\rm G}$-independent above this temperature.

\textbf{Temperature dependence of resistivity anisotropy} 
Figure~4 shows the in-plane anisotropy of the surface resistivity $\rho _{c}/\rho _{a}$ up to 200 K.
Here, $\rho _{c}$ ($\rho _{a}$) denotes the surface resistivity along the $c$-axis ($a$-axis) [the short axis of the elliptical FS is parallel to the $c$-axis as shown in Fig. 3(c)].
The color plots visually show the three domains, namely, the electron-doped, hole-doped, and Mott insulating states.
They are separated at low doping levels $|V_{\rm G}|\sim $0.1 V, which is in agreement with the occurrence of the BM behavior in Fig. 2.
The value of $\rho _{c}/\rho _{a}$ is highest in the hole-doped state, followed by the electron-doped (most isotropic), and the Mott insulating states. $\rho _{c}/\rho _{a}$ is less than 1 in the Mott insulating state as predicted by the conductivity calculations in our previous study \cite{Kawasugi2016}, although it is not clear why $\rho _{c}/\rho _{a}$ is much less than that in the electron-doped state.
As shown in the previous report \cite{Kawasugi2016}, the increase in $\rho _{c}/\rho _{a}$ under hole doping is attributable to the pseudogap formation that suppresses the carrier conduction along the $c$-axis.
At any gate voltage, the resistivity becomes more isotropic with increasing temperature, probably because the pseudogap diminishes due to thermal fluctuation at high temperatures, as implied by the temperature dependence of $R_{\rm Hs}$ and the CPT calculations.
However, the in-plane anisotropy in the hole-doped side remained even at 200 K.
All of the measurements are briefly summarized in Table II and the summarizing phase diagram is shown in Fig. 5.

\textbf{Discussion} 
Recently, we observed ambipolar superconductivity in a strain-tuned organic Mott EDLT based on $\kappa$-Cl \cite{Kawasugi2018}. 
The results described here have been measured in the same device with slightly larger tensile strain, or $U/t$.
Therefore, the BM and non-Fermi liquid behaviors observed in the transport properties of sample \#1 and \#2 are in close vicinity to the superconducting states. 
This type of bandfilling control experiments with tunable $U/t$ in the proximity of superconducting phase have become possible for the first time, by utilizing flexible organic lattice with bendable substrate. The details of the phase diagram such as existence/absence of possible quantum critical point will become possible in future works, although the present measurement points are not enough finely tuned because of the limitation of EDLT.

The BM behavior is suggested by various numerical studies \cite{Dobro2015,Deng2013,Xu2013,Kokalj2017}. 
The DMFT calculations \cite{Dobro2015}, for example, show that the BM behavior appears as a result of the Mott quantum criticality in the crossover region. 
Besides such deductive rationalization, an inductive extraction of BM behavior from experiments with different materials is also important \cite{Tyler1998,Qazilbash2006,Limelette2003}, because extrinsic factors not related to pure Mott insulator should be excluded. 
Our present results have clearly confirmed the emergence of the BM behavior in the doped organic Mott insulator, $\kappa$-BEDT-TTF EDLT.

According to the Hall measurements, the hole-doped $\kappa$-Cl resembles lightly or optimally hole-doped cuprates such as La$_{2-x}$Sr$_{x}$CuO$_{4}$ \cite{Ando2004}.
The strongly temperature-dependent $R_{\rm Hs}$ (which appears proportional to 1/$T$ at high temperatures) and the quadratic-in-temperature Hall angle in $\kappa$-Cl are also in common with the so-called strange-metal phase in the hole-doped cuprates.
This analogy supports the presence of pseudogap in the hole-doped $\kappa$-Cl.

On the other hand, the electron-doped $\kappa$-Cl appears quite different from the electron-doped cuprates. 
In particular, the value and even the sign of $R_{\rm H}$ in the electron-doped cuprates are generally temperature-dependent and are uncorrelated to the volume of the noninteracting hole-like Fermi surface \cite{Armitage2010}. 
This is contrasting to the electron-doped $\kappa$-Cl that shows almost temperature independent $R_{\rm Hs}$ simply corresponding to the noninteracting Fermi surface with hole density 1$-\delta$. 
Its transport properties are rather similar to those of the over-hole-doped cuprates such as La$_{2-x}$Sr$_{x}$CuO$_{4}$ \cite{Takagi1992,Ando2004} and Tl$_{2}$Ba$_{2}$CuO$_{6+\delta}$ \cite{Mackenzie1996} where the resistivity is proportional to $\sim T^{\alpha}$ with $\alpha$ between 1 and 2, $R_{\rm H}$ corresponds to the large holelike Fermi surface, and the Hall angle vs $T^{2}$ plots bend downward.

The similarities in the hole-doped states and differences in the electron-doped states between $\kappa$-Cl and the cuprates indicate the difference of their doping asymmetry.
In $\kappa$-Cl, electron doping induces more drastic transitions in resistivity compared to hole doping.
Similar doping asymmetry has also been observed for the superconducting region in the strain-controlled EDLT \cite{Kawasugi2018}. 
Unlike the hole-doped superconductivity, the electron-doped superconductivity shows very abrupt and discontinuous (first-order-like) transition.
Although the first-order transitions are not observed in this study, they may appear at extremely low temperatures because of large $U/t$ as shown in the DMFT calculations \cite{Dobro2015}.
This asymmetry may be attributable to the difference of their noninteracting band structures and resultant single-particle spectral functions in the Mott insulating state, where $t'$ exists only for one diagonal ($\kappa$-Cl), or for both diagonals (cuprates) of the sites.

To summarize, we investigated the surface transport properties of the EDLT based on an organic Mott insulator $\kappa$-Cl, which forms an anisotropic triangular lattice and has vHs below the Fermi level (hole-doped side) in the noninteracting band structure. 
In the hole-doped state, non-Fermi-liquid behaviors such as the BM behavior, strongly temperature-dependent $R_{\rm Hs}$, and the quadratic-in-temperature Hall angle are observed similarly to the strange metal phase in the hole-doped cuprates.
On the other hand, the doping asymmetry in $\kappa$-Cl appears different from that in the cuprates: the transport properties of electron-doped $\kappa$-Cl are similar to heavily hole-doped cuprates, and not likely to be related to those of the electron-doped cuprates.
The electron-doped $\kappa$-Cl shows more Fermi-liquid-like behaviors such as the quadratic-in-temperature resistivity at low temperatures and the temperature-independent $R_{\rm H}$ nearly corresponding to the large noninteracting Fermi surface.
The difference of the doping asymmetry between $\kappa$-Cl and the cuprates may originate from the difference of their band structures. 
At high temperatures, the BM behavior widely appears in $\kappa$-Cl regardless of the doping polarity, supporting the universality of the BM transport near the Mott transitions.
Even in the sufficiently electron-doped state, which shows Fermi-liquid behavior at low temperatures, the system still shows the BM behavior at high temperatures probably because of the proximity effects of the Mott quantum criticality.

Finally, we emphasize that determination of physical properties in a wide range of thermodynamic parameters in a single sample is quite important in understanding Mott physics. 
Because of its low half-filling carrier density, $\kappa$-Cl is one of the most appropriate materials to study Mott physics in wide filling control using field effect without electrical breakdown or electrochemical degradation.
We have found that the BM behavior at high temperatures persists in almost all doping levels, while the pseudogap behavior at low temperatures is quite band-structure dependent, implying the different energy scales and mechanisms governing these two behaviors.

\clearpage
\begin{figure}[htbp]
  \begin{center}
    \includegraphics[width=8.0cm]{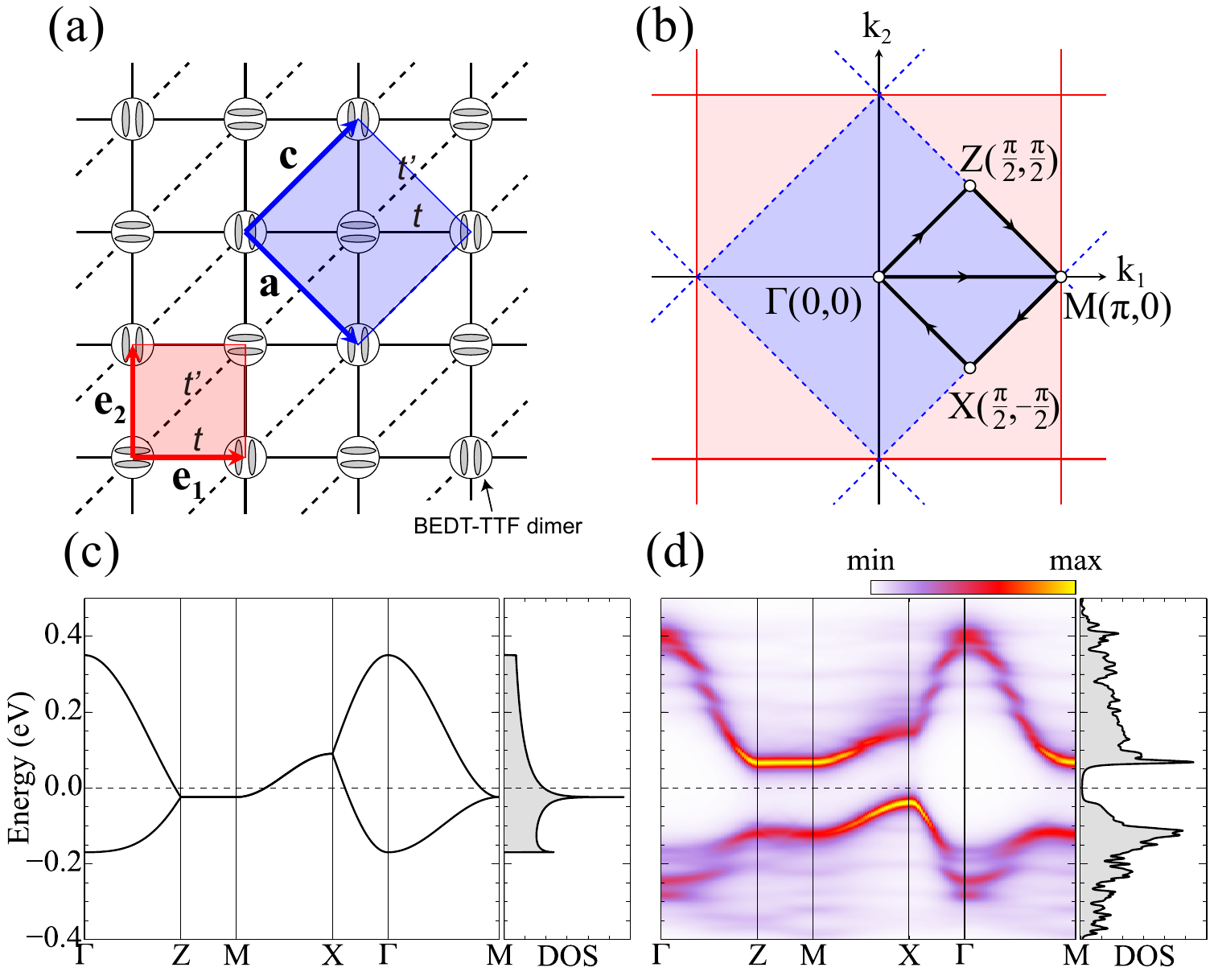}
  \caption{Unit cells, Brillouin zones, band structure, and single-particle spectral function of $\kappa$-Cl. Note that the calculations are based on the one-band model. However, the band structure and spectral functions are shown in the two-site Brillouin zone [blue shaded area in (b)] because the adjacent BEDT-TTF dimers are not completely equivalent in the material. Accordingly, the X, Z, and M points in (c), (d) correspond to the points ($\pi/2$, $-\pi/2$), ($\pi/2$, $\pi/2$), and ($\pi$, 0) in the Brillouin zone of the one-site unit cell.
  (a) Schematic of the anisotropic triangular lattice of $\kappa$-Cl. The translational vectors $\rm{\bf{e_{1}}}$ and $\rm{\bf{e_{2}}}$ ($\bf{a}$ and $\bf{c}$) are represented by the red (blue) arrows. The red (blue) shaded region represents the unit cell containing one site (two sites). The ellipses on the sites denote the conducting BEDT-TTF dimers. 
  (b) The momentum space for the anisotropic triangular lattice. The Brillouin zones of the one- (two)-site unit cell is represented by the red (blue) shaded region bounded by the red solid (blue dashed) lines.
  (c) Noninteracting tight-binding band structure along high symmetric momenta and density of states (DOS) of $\kappa$-Cl ($t'/t$ = $-$0.44 with $t$ = 65 meV). The Fermi level for half filling is set to zero and denoted by dashed lines. (d) Single-particle spectral functions and DOS of $\kappa$-Cl at half filling in an antiferromagnetic state at zero temperature, calculated by variational cluster approximation \cite{Kawasugi2018}. The Fermi level is denoted by dashed lines at zero energy. The momentum path along $\Gamma$-Z-M-X-$\Gamma$-M used in (c) and (d) is indicated by arrows in (b).}
  \end{center}
\end{figure}

\begin{figure}[htbp]
  \begin{center}
    \includegraphics{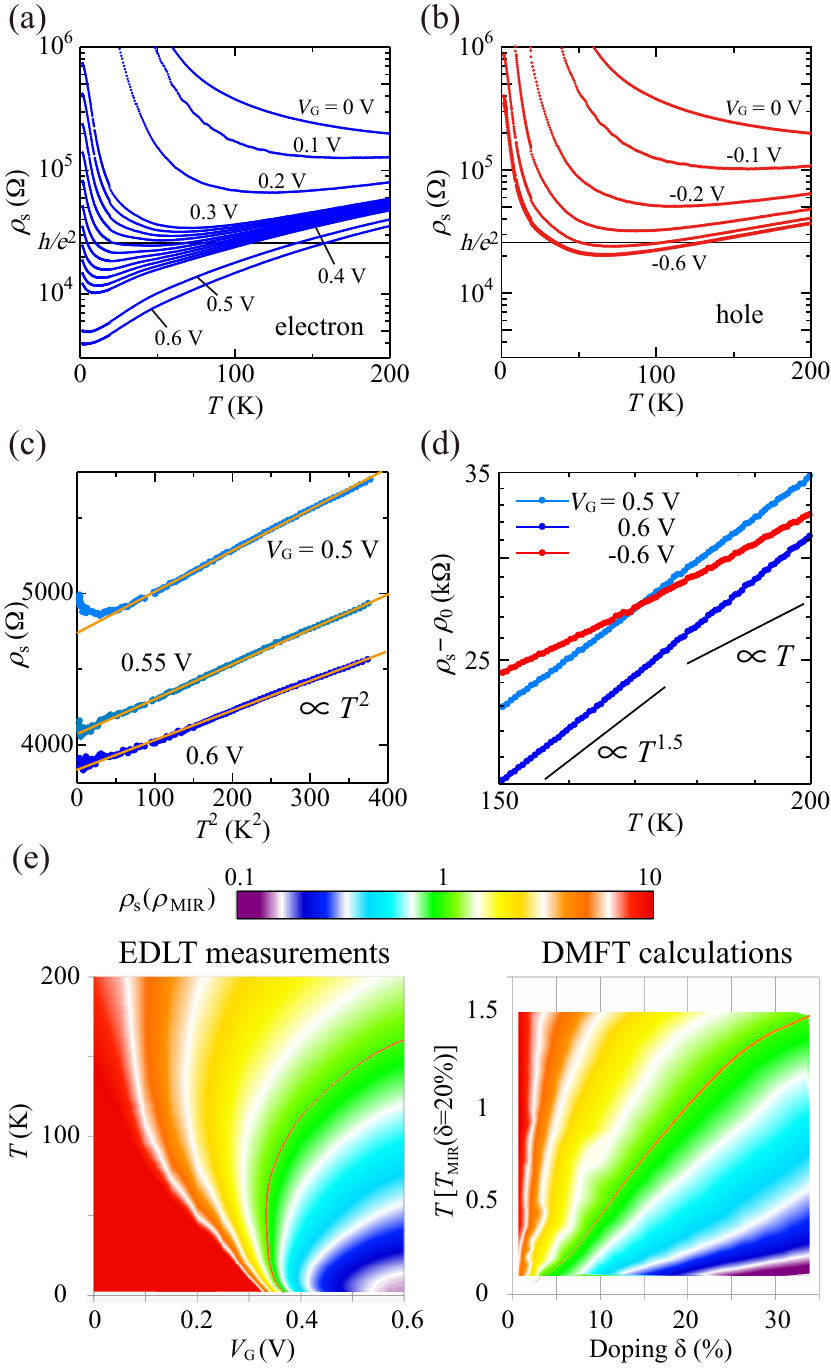}
  \caption{Temperature ($T$) dependence of surface resistivity $\rho _{\rm s}$ along $a$-axis in sample \#1 under (a) electron doping and (b) hole doping. (c) $\rho _{\rm s}$ vs $T^{2}$ plots below 20 K for $V_{\rm G}$ = +0.5, +0.55, and +0.6 V. The orange solid lines denote $\rho _{\rm s} \propto T^{2}$ lines. (d) Log-log plots of $\rho _{\rm s}-\rho _{\rm 0}$ vs $T$ for $V_{\rm G}$ = +0.5 and +0.6 V. The plot for $-$0.6 V is also shown as a reference, where $\rho _{\rm 0}$ at +0.6 V is assumed. (e) Color plots of surface resistivity under electron doping. The original data (52 temperature cycles at different $V_{\rm G}$ values) are shown in (see Fig. S3 in the Supplemental Material \cite{SupplementalMaterial}). The orange solid line indicates the contour line $\rho _{\rm MIR}$. Right panel shows the results of DMFT calculations in Ref. \cite{Dobro2015}, where the temperature is normalized with $T_{\rm MIR}$, temperature at which the resistivity reaches $\rho_{\rm MIR}$ for $\delta=20$\%. }
  \end{center}
\end{figure}

\renewcommand{\arraystretch}{1.5}
\begin{table*}[htbp]
\caption{Properties of doped Mott insulators.}
\begin{center}
\begin{tabular}{c||c|c||c|c|}
 & \multicolumn{2}{c||}{High-$T_{\rm C}$ cuprates} & \multicolumn{2}{c|}{$\kappa$-type BEDT-TTF salts}\\
\cline{2-5}
 & Electron doping & Hole doping & Electron doping & Hole doping\\
\hline
\hline
Lattice & \multicolumn{2}{c||}{Square} & \multicolumn{2}{c|}{Triangular} \\
\hline
$t'/t$ & \multicolumn{2}{c||}{$<0$} & \multicolumn{2}{c|}{$<0$} \\
\hline
Orbital & Cu(3d) & O(2p) & Upper HOMO & Upper HOMO \\
\hline
Superconductivity & Yes & Yes & Yes \cite{Kawasugi2018} & Yes \cite{Kawasugi2018,Taniguchi2007,Oike2015} \\ 
\hline
Pseudogap & Yes & Yes & No \cite{Kawasugi2016} & Yes \cite{Kawasugi2016} \\ 
\hline
Bad metal behavior & Yes & Yes & Yes [This work] & Yes \cite{Taniguchi2007,Oike2015}, [This work] \\
\hline
\end{tabular}
\end{center}
\end{table*}

\begin{figure}[htbp]
  \begin{center}
    \includegraphics{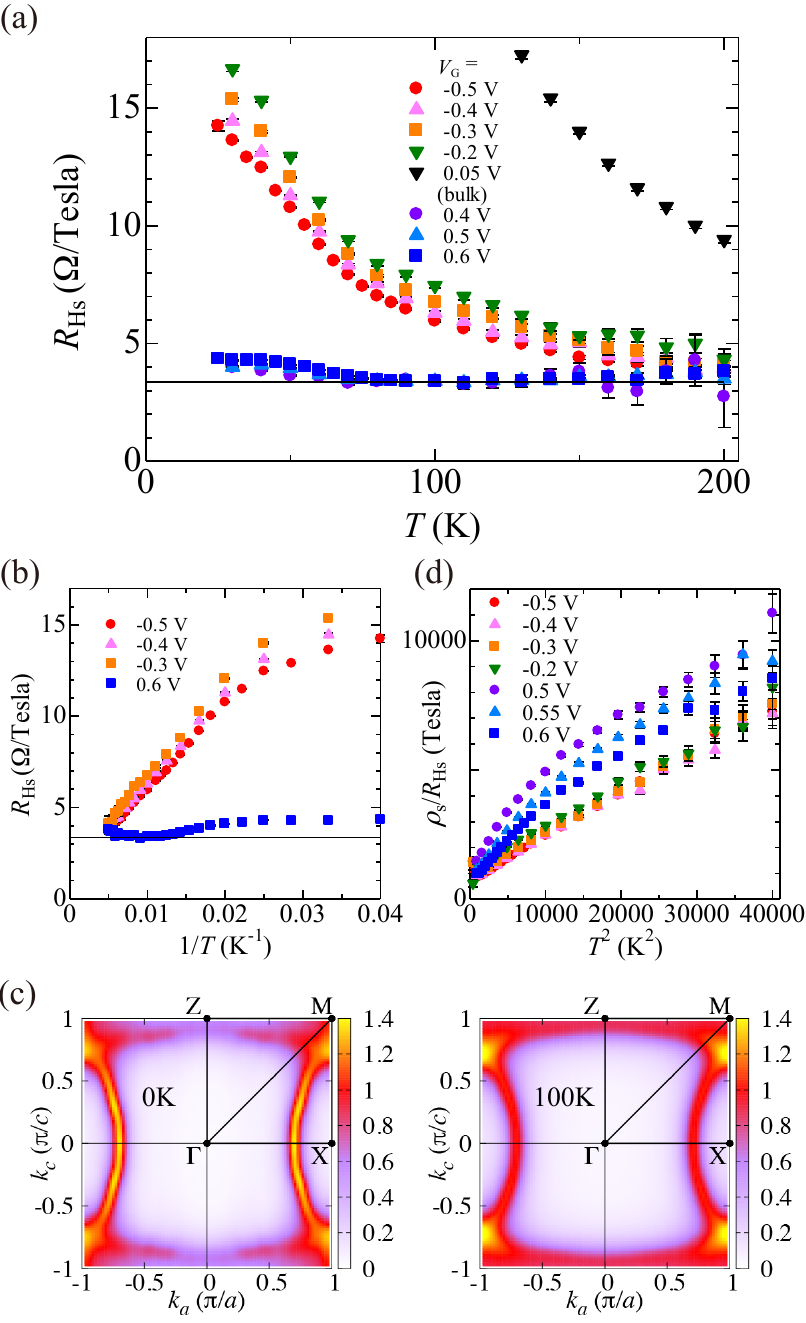}
  \caption{Hall measurements in sample \#2. (a) Temperature dependence of $R_{\rm Hs}$. The dashed line indicates 1/$en_{\rm HF}$, where $n_{\rm HF}$ is the hole density per unit cell at the charge neutrality point. (b) Same as (a) but $R_{\rm Hs}$ vs 1/$T$ plots. (c) CPT calculations of the Fermi surface for 17\% hole doping at 0 K (left) and 100 K (right). (d) Hall angle at 1 Tesla vs $T^{2}$ plots.
  }
  \end{center}
\end{figure}

\begin{figure}[htbp]
  \begin{center}
    \includegraphics{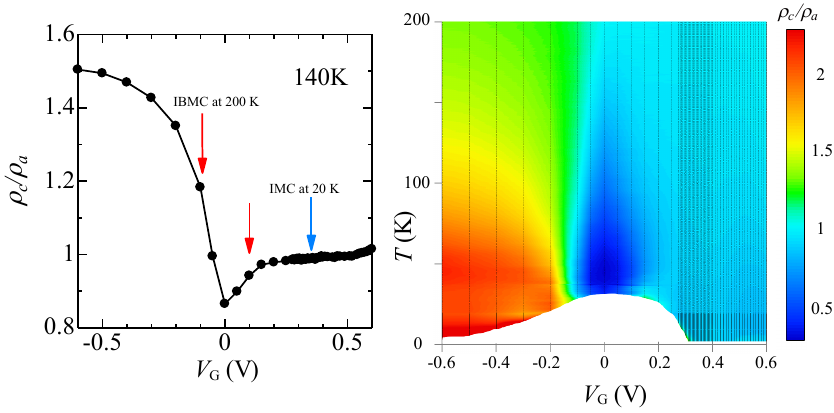}
  \caption{Temperature dependence of in-plane anisotropy of surface resistivity in sample \#1 (Note that both $a$- and $c$-axes are parallel to the conducting plane in this material). The black dots denote the data points in both panels. The left panel shows the height profile at 140 K where IBMC (indicated by red arrows) and IMC (blue arrow) denote insulator-to-BM crossover and insulator-to-metal crossover, respectively. Data are missing at low temperatures and low doping (white region in the right panel) due to the high resistance.}
  \end{center}
\end{figure}

\renewcommand{\arraystretch}{1.5}
\begin{table*}[htbp]
\caption{Transport properties of the $\kappa$-Cl EDLT in this work. The power-law exponents of the BM behaviors are roughly estimated values assuming constant cell parameters.}
\begin{center}
\begin{tabular}{c||c|c||c|c|}

 & \multicolumn{2}{c||}{Electron doping} & \multicolumn{2}{c|}{Hole doping}\\
\cline{2-5}
 & High $T$ & Low $T$ & High $T$ & Low $T$\\
\hline
\hline
Resistivity & Bad metal ($\propto T^{1.5}$) & Fermi liquid ($\propto T^{2}$) & Bad metal ($\propto T$) & Semiconductor \\
\hline
Hole density per site & 1$-\delta$ & 1$-\delta$ & Activated up to 1$-\delta$ & $\sim$(1$-\delta$)/3 \\ 
\hline
Hall angle & Non-quadratic & $\propto T^{2}$ & \multicolumn{2}{c|}{$\propto T^{2}$} \\ 
\hline
Fermi surface (CPT) & Elliptical & Elliptical & Elliptical & Lens-like (pseudogap) \\
\hline
Resistivity anisotropy & Low & Low & Medium & High \\
\hline
Superconductivity  & \multicolumn{2}{c||}{Yes (first-order like, narrow $V_{\rm G}$ region) \cite{Kawasugi2018}} & \multicolumn{2}{c|}{Yes (wide $V_{\rm G}$ region) \cite{Kawasugi2018}} \\
\hline
\end{tabular}
\end{center}
\end{table*}

\begin{figure}[htbp]
  \begin{center}
    \includegraphics{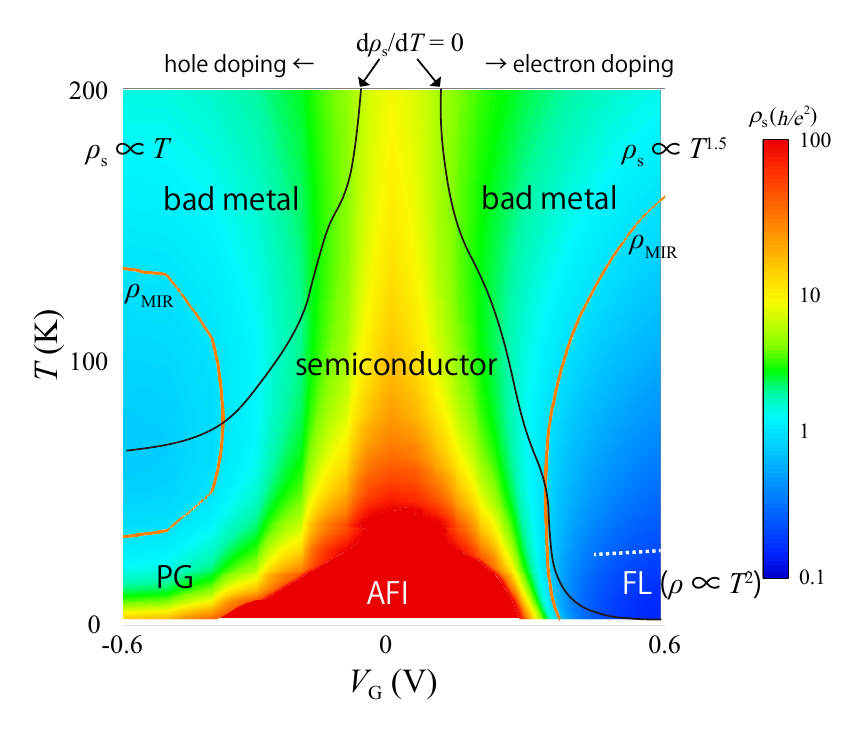}
  \caption{Phase diagram drawn from the transport properties of sample \#1. PG, AFI, and FL denote pseudogap, antiferromagnetic insulator, and Fermi liquid, respectively. Black lines indicate temperatures where d$\rho _{\rm s}$/d$T=0$. Above this temperature, the conductivity is metallic-like. Orange lines indicate temperatures and $V_{\rm G}$  where $\rho _{\rm s}=\rho _{\rm MIR}$. The surface resistivity is quadratic in temperature below the white dashed line in the electron-doped side. The surface resistivity $\rho _{\rm s}$ is shown as a color plot.}
  \end{center}
\end{figure}

\section*{Acknowledgments}
We thank Dr. Vladimir Dobrosavljevic for discussion. We would like to acknowledge Teijin DuPont Films Japan Limited for providing the PET films.
This work was supported by MEXT and JSPS KAKENHI (Grant Nos. JP16H06346, 15K17714, 26102012 and 25000003), JST ERATO, MEXT Nanotechnology Platform Program (Molecule and Material Synthesis), 
and HPCI System Research Project (Project Nos. hp160122 and hp170324). 
We are also grateful for allocating computational time of the HOKUSAI GreatWave and HOKUSAI BigWaterfall 
supercomputer at RIKEN Advanced Center for Computing and Communication (ACCC), and 
the K computer at RIKEN Center for Computational Science (R-CCS). 
K.S. acknowledges support from the JSPS Overseas Research Fellowships.

\end{document}